\documentclass[sigconf]{acmart}

\copyrightyear{2024}
\acmYear{2024}
\setcopyright{acmlicensed}
\acmConference[LAMPS '24] {Proceedings of the 1st ACM Workshop on Large AI Systems and Models with Privacy and Safety Analysis}{October 14--18, 2024}{Salt Lake City, UT, USA.}
\acmBooktitle{Proceedings of the 1st ACM Workshop on Large AI Systems and Models with Privacy and Safety Analysis (LAMPS '24), October 14--18, 2024, Salt Lake City, UT, USA}
\acmISBN{979-8-4007-1209-8/24/10}
\acmDOI{10.1145/3689217.3690617}

\usepackage{booktabs} 
\usepackage{subfigure}
\usepackage{multirow}
\usepackage{enumitem}
\usepackage{pifont}
\usepackage{xurl}
\usepackage{balance}
\usepackage{mathtools}
\usepackage{amsfonts}

\usepackage[linesnumbered,ruled,vlined]{algorithm2e}
\usepackage{graphicx}
\usepackage{textcomp}
\usepackage{colortbl}
\usepackage{xcolor}
\usepackage{caption}

\usepackage[prefixes-as-symbols=false]{siunitx}

\usepackage{mdframed} 
\usepackage{lipsum}  
\usepackage{makecell}

\begin{document}

\title{CryptoTrain: Fast Secure Training on Encrypted Dataset}

\author{Jiaqi Xue}
\email{jiaqi.xue@ucf.edu}
\affiliation{
  \institution{University of Central Florida}
  \city{Orlando}
  \state{Florida}
  \country{USA}
}
\author{Yancheng Zhang}
\email{yancheng.zhang@ucf.edu}
\affiliation{
  \institution{University of Central Florida}
  \city{Orlando}
  \state{Florida}
  \country{USA}
}
\author{Yanshan Wang}
\email{yanshan.wang@pitt.edu}
\affiliation{
  \institution{University of Pittsburgh}
  \city{Pittsburgh}
  \state{Pennsylvania}
  \country{USA}
}
\author{Xueqiang Wang}
\email{xueqiang.wang@ucf.edu}
\affiliation{
  \institution{University of Central Florida}
  \city{Orlando}
  \state{Florida}
  \country{USA}
}
\author{Hao Zheng}
\email{hao.zheng@ucf.edu}
\affiliation{
  \institution{University of Central Florida}
  \city{Orlando}
  \state{Florida}
  \country{USA}
}
\author{Qian Lou}
\email{qian.lou@ucf.edu}
\affiliation{
  \institution{University of Central Florida}
  \city{Orlando}
  \state{Florida}
  \country{USA}
}


\renewcommand{\shortauthors}{Jiaqi Xue et al.}

\begin{abstract}

Secure training, while protecting the confidentiality of both data and model weights, typically incurs significant training overhead. Traditional Fully Homomorphic Encryption (FHE)-based non-inter-active training models are heavily burdened by computationally demanding bootstrapping. To develop an efficient secure training system, we established a foundational framework, CryptoTrain-B, utilizing a hybrid cryptographic protocol that merges FHE with Oblivious Transfer (OT) for handling linear and non-linear operations, respectively. This integration eliminates the need for costly bootstrapping. Although CryptoTrain-B sets a new baseline in performance, reducing its training overhead remains essential. We found that ciphertext-ciphertext multiplication (CCMul) is a critical bottleneck in operations involving encrypted inputs and models. Our solution, the CCMul-Precompute technique, involves precomputing CCMul offline and resorting to the less resource-intensive ciphertext-plaintext multiplication (CPMul) during private training. Furthermore, conventional polynomial convolution in FHE systems tends to encode irrelevant and redundant values into polynomial slots, necessitating additional polynomials and ciphertexts for input representation and leading to extra multiplications. Addressing this, we introduce correlated polynomial convolution, which encodes only related input values into polynomials, thus drastically reducing the number of computations and overheads. By integrating CCMul-Precompute and correlated polynomial convolution into CryptoTrain-B, we facilitate a rapid and efficient secure training framework, CryptoTrain. Extensive experiments demonstrate that CryptoTrain achieves a $\sim5.3 \times$ training time reduction compared to prior methods. 

\end{abstract}

\keywords{Secure Training, Fully Homomorphic Encryption, Oblivious Transfer, Data Privacy, Machine Learning}

\ccsdesc[500]{Security and privacy~Human and societal aspects of security and privacy}
\ccsdesc[500]{Computing methodologies~Machine learning}

\keywords{Secure training; Homomorphic encryption; Oblivious transfer}

\maketitle
\section{Introduction}
\label{t:intro}

Training as a Service (TaaS)~\cite{mohassel2017secureml} is crucial in today's dynamic digital landscape, providing scalable, flexible, and uniform training across various fields. Big companies based on powerful computational resources, such as AWS~\cite{Amazon}, Google Cloud~\cite{Google}, and Azure~\cite{Mircrosoft}, are providing different TaaS for regular users. It's employed in various scenarios. For instance, a hospital may want to use patients' private information to train an AI-assisted diagnostic system, and in financial analysis, a company may wish to train an AI-assistant economic advisory system on its private business data. These scenarios involve the need to utilize private data for training models but encounter difficulties due to a lack of expertise and computational resources, leading them to seek the assistance of TaaS. Yet, as TaaS expands, so does the urgency to address data privacy. Stringent legal requirements and Federal regulations such as CCPA~\cite{CCPA} and GDPR~\cite{GDPR} also underscore the need to ensure data and model privacy.


To address the mentioned concerns, several cryptographic techniques have been developed, albeit with distinct limitations. The FHE-only methods~\cite{nandakumar2019towards, lou2020glyph, lee2023hetal,zheng2024ofhe,lou2023vfhe} allow for computations on encrypted data, protecting privacy throughout the process on the server. Regrettably, non-interactive FHE-based training suffers from significant latency due to its deep circuit depth and costly FHE bootstrapping. Alternatively, MPC-based training~\cite{keller2022secure} divides the computational data among multiple parties, ensuring no single entity has full access to sensitive data. Despite enhancing privacy, the effectiveness of MPC is often hindered by the need for multiple non-colluding servers~\cite{cramer2015secure, keller2022secure, mohassel2017secureml}. 

To overcome these challenges, we first develop our baseline protocol, CryptoTrain-B, which strategically combines Oblivious Transfer (OT)~\cite{rathee2020cryptflow2, rathee2021sirnn} for non-linear computations and HE-based protocols~\cite{huang2022cheetah} for linear computations, as shown in Figure~\ref{fig:overview} where data and model are both encrypted during training.  CryptoTrain-B significantly reduces training latency by removing the need for costly FHE bootstrapping compared to FHE-only methods~\cite{nandakumar2019towards, lou2020glyph}. It also simplifies the threat model to a two-party system, involving just the communication between the client and a single server, which removes the demand for non-colluding servers. CryptoTrain-B achieves state-of-the-art secure training, but it still suffers from a large training time. To enhance the training speed of CryptoTrain-B, we have pinpointed two key issues and associated challenges:

\begin{figure}[t!]
    \centering
    \includegraphics[width=0.85\linewidth]{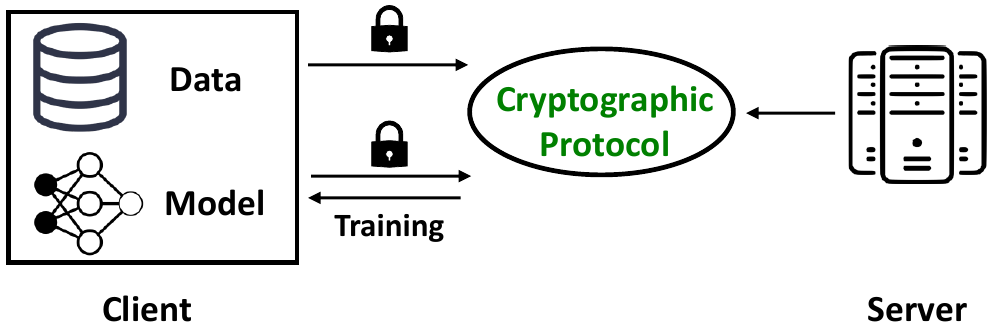}
           \vspace{-0.15in}
    \caption{Overview of secure training based on cryptographic protocols, e.g., FHE and OT. The lock represents that data and model are encrypted.}
       \vspace{-0.1in}
    \label{fig:overview}
\end{figure}

\begin{itemize}[leftmargin=*, nosep, topsep=0pt, partopsep=0pt]
    \item \textbf{Expensive Ciphertext-Ciphertext multiplication (CCMul)}: Present FHE-based linear operations, such as convolution and matrix multiplication~\cite{huang2022cheetah,han2022coxhe,yudha2024boostcom}, do not require costly rotations.  As opposed to secure inference where only the inputs are encrypted, both the model and inputs are encrypted in training. This makes CCMul the primary computational bottleneck, consuming over 90\% of the training time.
    
    
    \item \textbf{Correlation-unaware Polynomial Convolution}: The existing approach to polynomial convolution encodes inputs based on row-major order, and it even encodes unassociated zeros or paddings into the polynomial slots. As a result, an input often requires multiple polynomials for representation, leading to an increase in ciphertext multiplications. To address this, a new form of polynomial convolution is necessary, one that encodes correlated values into the polynomial to maximize the utilization of polynomial slots and minimize the number of ciphertext multiplications.

\end{itemize}

To enhance the efficiency of CryptoTrain-B and address the outlined challenges, we introduce two innovative techniques: \textit{CCMul-Precompute} and \textit{Correlation-aware Polynomial Convolution}, which collectively contribute to the development of a faster secure training framework, CryptoTrain. Specifically, \textit{CCMul-Precompute} is designed to carry out the computationally intensive CCMul operations in an offline mode, enabling the use of less resource-intensive ciphertext-plaintext multiplication (CPMul) during actual training, thereby accelerating the secure training process. Meanwhile, \textit{Correlation-aware Polynomial Convolution} focuses on encoding only the correlated input values into polynomial slots, facilitating more effective packing and substantially decreasing the number of ciphertext multiplications required. CryptoTrain demonstrates a significant reduction in training time, achieving a $5.3\times$ speed improvement over state-of-the-art method~\cite{keller2022secure} on the CIFAR-10 dataset. Moreover, this study represents the first attempt at secure training on the TinyImageNet dataset. The results validate the effectiveness of the two techniques we proposed, showcasing an impressive speedup of up to $8.2\times$ compared to the established baseline CryptoTrain-B.

\section{Background and Motivation}
\label{s:background}

\noindent\textbf{Threat Model.} Our threat model, as illustrated in Figure~\ref{fig:overview}, considers servers that are semi-honest, where a cloud server follows the cryptographic protocol yet may attempt to infer information from the client's dataset and model weights. Under this threat model, the client aims to leverage the server's computational resources to train a neural network model on their dataset privately. The client (data owner), encrypts their data before sending it to the server. The server works with the encrypted data and sends back the encrypted model weights. Since only the client holds the private key, they alone can decrypt the model weights. 

\begin{figure}[h!]
   \centering
   \includegraphics[width=\linewidth]{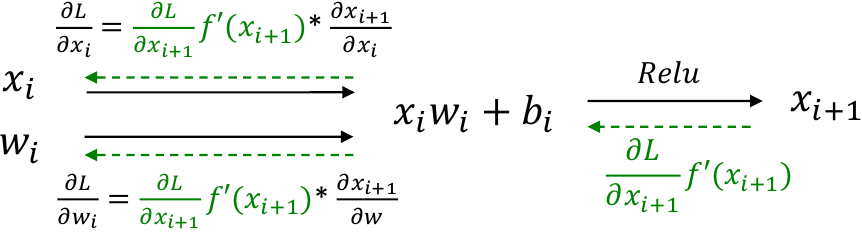}
   \caption{Forward and backward propagations.}
   \label{fig:Backward}
\end{figure}

\noindent \textbf{Neural Network Training.}
Neural network training is an iterative process comprising forward and backward propagation. During forward propagation, input data sequentially traverses through the layers. As shown in Figure~\ref{fig:Backward}, forward propagation for a fully connected layer can be mathematically expressed as $x_{i+1}=f(x_iw_i+b_i)$, where $x_i$ denotes the input to the $i$-th layer, $w_i$ and $b_i$ represent the respective weights and bias of the layer, and $f(\cdot)$ is the activation function. Conversely, backward propagation, which is responsible for error correction and model updating, can be characterized by the gradients $\frac{\partial L}{\partial x_i}=\frac{\partial L}{\partial x_{i+1}}f'(x_{i+1})\frac{\partial x_{i+1}}{x_i}$, $\frac{\partial L}{\partial w_i}=\frac{\partial L}{\partial x_{i+1}}f'(x_{i+1})\frac{\partial x_{i+1}}{\partial w}$ and $\frac{\partial L}{\partial b_i}=\frac{\partial L}{\partial x_{i+1}}f'(x_{i+1})$, with $\frac{\partial x_{i+1}}{\partial w_i}= x_i$ and $\frac{\partial x_{i+1}}{\partial x_i}=w_{i}^T$. $L$ denotes the loss. Moreover, for convolution layer, the backward propagation algorithm utilizes convolution operations to calculate gradients~\cite{maheshwari2021backpropagation, kafunah2016backpropagation, dumoulin2016guide}. Specifically, the gradient of the kernel $w_i$ is obtained by convolving the input $x_i$ with the gradient of output $\frac{\partial L}{\partial x_{i+1}}$, while the gradient of $x_i$ is derived by convolving the $w_i$ with $\frac{\partial L}{\partial x_{i+1}}$. The whole training process, encompassing both forward and backward propagation, requires substantial padding to maintain consistent feature map dimensions and ensure the alignment of gradient tensors~\cite{dumoulin2016guide}.

\noindent\textbf{Secure Training. }
Secure training involves methods and protocols aimed at safeguarding the confidentiality of data and models during the training phase of machine learning algorithms, particularly when this process is delegated to external servers. This approach guarantees the protection of sensitive information, such as proprietary datasets and model weights. Current research in secure training predominantly concentrates on developing cryptographic methods to enable computations that preserve privacy. A prominent method involves FHE, enabling computations on encrypted data~\cite{lou2020falcon,nandakumar2019towards, lou2020autoprivacy}. In FHE-based approaches, the process typically unfolds as follows: the client encrypts their data and sends it to the server; the server processes this encrypted data; and finally, the server returns encrypted results to the client for decryption. While this approach necessitates client-side encryption and decryption, modern hardware optimizations~\cite{samardzic2021f1, kim2023sharp, aikata2023reed, zheng2023priml} have made these operations more efficient. Moreover, the encryption process remains consistent regardless of the server-side operations, e.g., convolution~\cite{huang2022cheetah} or attention~\cite{zheng2023primer}, simplifying the client's computational considerations.
However, such non-interactive techniques face substantial delays due to deep circuit depths and the intensive bootstrapping requirements of FHE. For example, Glyph~\cite{lou2020glyph} took approximately 8 days to train a network on the MNIST dataset. Most recently, HETAL~\cite{lee2023hetal} reduces the HE-based private training time by adopting transfer learning and early stopping, yet it does not involve optimizing the ciphertext's computations.

An alternative approach is Multi-Party Computation (MPC), where data and models are distributed across multiple entities, with computations structured so no single party has complete access to all information. However, MPC-based secure training typically depends on several non-colluding servers for effectiveness. For instance, previous studies~\cite{mohassel2017secureml,agrawal2019quotient,keller2022secure,rathee2023secure, jawalkar2023orca} require at least two servers, while others~\cite{mohassel2018aby3,wagh2019securenn,wagh2020falcon,attrapadung2021adam} need three, and Trident~\cite{chaudhari2019trident} even four. The necessity for multiple servers introduces practical issues, particularly concerning the assurance of non-collusion. This paper, however, focuses on a single-server secure training protocol, thereby simplifying the threat model and enhancing its practical application without sacrificing security.


\noindent \textbf{Cryptographic Preliminaries. }
Our protocols synergistically combine Additive Secret Sharing (ASS), Homomorphic Encryption (HE) for private linear operations, and Oblivious Transfer (OT) for private non-linear computations, leveraging the unique strengths of each to ensure robust security and efficiency.

\begin{itemize}[leftmargin=*, nosep, topsep=0pt, partopsep=0pt]
\item \textbf{ASS. }We leverage a 2-out-of-2 ASS scheme~\cite{cramer2015secure}, over the ring $\mathbb{Z}_{\ell}$, where $\ell$ is the bitwidth of input $x$. This scheme decomposes 
$x$ into two random shares, $\left \langle x \right \rangle_{0}$ and $\left \langle x \right \rangle_{1}$, satisfying $x = (\left \langle x \right \rangle_{0}+\left \langle x \right \rangle_{1})$ mod $ \mathbb{Z}_{\ell}$. $P_0$ and $P_1$ hold $\left \langle x \right \rangle_{0}, \left \langle x \right \rangle_{1}$, respectively. As established by~\cite{cramer2015secure}, this configuration ensures that neither party can ascertain the actual value of $x$. Importantly, ASS inherently supports linear computations without the need for inter-party communication, such as addition and multiplication by a constant.

\item \textbf{HE. } HE~\cite{elgamal1985public, paillier1999public, lou2019she} is a widely used cryptographic technique 
in both private training and inference~\cite{lou2021hemet, lou2021safenet, xue2022audit}. We adopt HE to enable linearly homomorphic operations on ciphertext. A HE cryptosystem is designed to encrypt a plaintext $p$ into a ciphertext $c$ using a function $\epsilon(p,pk)$, with public key $pk$. The decryption function $\sigma$ reverts the ciphertext $c$ to its plaintext form $p$, formulated as $p=\sigma(c,sk)$, with private key $sk$. An operation $
\otimes$ is \textit{homomorphic}, if there is another operation $\circ$ satisfying the relation: $\sigma(\epsilon(x,pk) \circ \epsilon(y, pk), sk)=\sigma(\epsilon(x\otimes y,pk),sk)$, where $x$ and $y$ are plaintext operands. 


\item \textbf{OT. } We use OT to execute non-linear computation, as detailed in works~\cite{rathee2020cryptflow2, rathee2021sirnn}. Specifically, we utilize the 1-out-of-2 correlated OT~\cite{asharov2013more} (2-COT$_{\ell}$) and 1-out-of-$k$~\cite{kolesnikov2013improved} ($k$-OT$_{\ell}$). In 2-COT$_{\ell}$, the sender inputs a correlation value $x \in \mathbb{Z}_{\ell}$ and the receiver inputs a bit $i\in\{0,1\}$. The protocol outputs a random element $r \in \mathbb{Z}_{\ell}$ to the sender and the value $r+i\cdot x$ to the receiver.
Conversely, in $k$-OT$_{\ell}$, the sender has $k$ distinct messages $m_0,...m_{k-1}$, and the receiver inputs an index $i \in [k]$. This protocol enables the receiver to learn message $x_i$ without gaining any knowledge about the other messages $x_j$ where $j\in [k]$ and $j\ne i$, while the sender learns nothing about the receiver's chosen index $i$.


\item \textbf{Notations.}
 For linear functions such as matrix multiplication and convolution, we pack matrices into polynomials~\cite{huang2022cheetah, mishra2021fast}. Let $\mathbb{A}_{N,p}$ denote the set of integer polynomials, $\mathbb{A}_{N,p} = \mathbb{Z}_p[x] / (x^N + 1)$, where $N$ is a 2-power number and $\mathbb{Z}_p$ is the set of coefficients defined on prime number $p>0$. We use  caret symbol over lower-case letters to denote a polynomial, e.g., $\hat{a}$, and the ciphertext form encrypted by HE is $[a]$.
 

\end{itemize} 

\begin{figure}[t!]
    \centering
    \includegraphics[width=\linewidth]{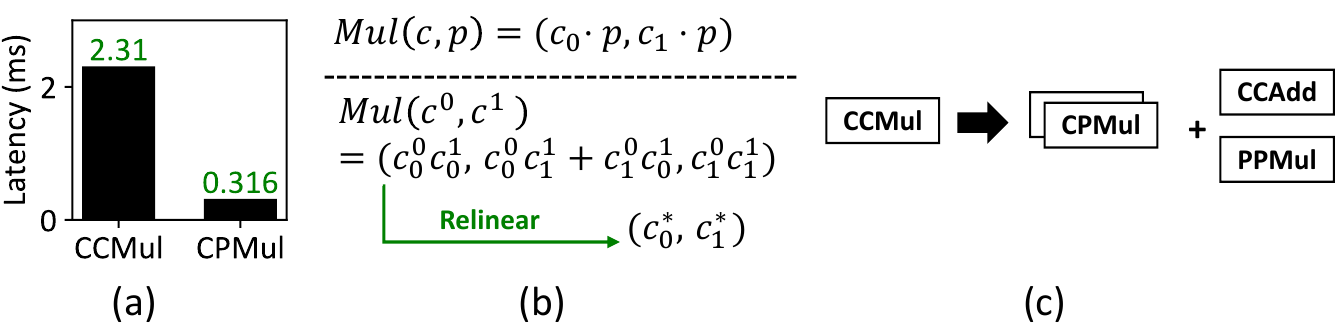}
    \caption{Motivation to precompute CCMul and use CPMul during online training. (a) CCMul is more expensive than CPMul. (b) CCMul needs expensive relinearization. (c) We replace a CCMul with 2 CPMul and negligible CCAdd and PPMul. }
    \vspace{-0.5cm}
    \label{fig:motivation}
\end{figure}

\noindent\textbf{Motivation.}
\label{s:motivation}
In secure training, HE-based linear operations such as convolutions typically account for about 96\% (training an AlexNet model on TinyImageNet) of the computation overhead, making them the primary bottleneck~\cite{huang2022cheetah, keller2022secure}. Therefore, our research focuses on optimizing the performance of these linear layers.
Figure~\ref{fig:motivation}(a) reveals that linear operation CCMul has a considerably higher latency, being about 7 times slower than CPMul. This delay in CCMul mainly stems from the need for relinearization, a process crucial for controlling the size of ciphertexts after multiplication. As illustrated in Figure~\ref{fig:motivation}(b), relinearization, which converts $(c_0^0c_0^1, c_0^0c_1^1+c_1^0c_0^1, c_1^0c_1^1)$ into $(c^0, c^1)$, entails intricate polynomial calculations, resulting in substantial computational overhead. On the other hand, CPMul naturally yields a two-component output, $(c_0p, c_1p)$, obviating the need for relinearization. This inherent advantage of CPMul has led us to devise a method that replaces CCMul with CPMul and other less demanding operations. Our strategy, as shown in Figure~\ref{fig:motivation}(c), includes ciphertext-ciphertext addition (CCAdd) and plaintext-plaintext multiplication (PPMul). This innovative approach aims to considerably diminish the computational burden of CCMul by capitalizing on the efficiency of CPMul for more efficient processing.

Padding plays a crucial role in CNN training during both forward and backward propagation, maintaining feature map dimensions. For instance, a $32\times32$ input might need padding to $36\times36$ for a $5\times5$ kernel to keep the output size consistent with the input. In back propagation, the demand for padded convolution escalates as it computes gradients for inputs $x_i$ and weights $w_i$. 
However, the current convolution protocols~\cite{huang2022cheetah}, which use polynomial multiplications for convolution, are inefficient with zero-padding. As depicted in Figure~\ref{fig:Padding}(a), unnecessary computations like $c_8x^8$ and $c_{12}x^{12}$ waste polynomial degrees without contributing to the output. This inefficiency often leads to dividing a single polynomial multiplication into four parts. To overcome this, we aim to develop a more efficient mapping scheme that eliminates these redundancies and reduces the number of required multiplication operations.
\section{CryptoTrain}
\label{s:method}
In this section, we present our cryptographic protocols for secure model training. As outlined in Section~\ref{s:background}, the training process comprises forward and backward propagation. Both of these can be decomposed into linear operations (e.g., convolution) and non-linear operations (e.g., activation functions). Our work primarily focuses on optimizing the linear layers, while for non-linear layers, we leverage state-of-the-art OT-based protocols~\cite{rathee2020cryptflow2, rathee2021sirnn}.

The structure of this section is as follows: Section~\ref{s:CC_baseline} introduces our baseline protocol CryptoTrain-B, overcoming the need for non-colluding servers. In Section~\ref{s:CP_cryptotrain}, we introduce CryptoTrain-P, enhancing efficiency by offloading CCMul to preprocessing phase. Lastly, Section~\ref{s:padding} describes our correlation-aware convolution protocol to improve efficiency in convolutions with padding. The culmination of these developments is CryptoTrain, a comprehensive protocol embodying these enhancements to secure training.

\subsection{CryptoTrain-B}
\label{s:CC_baseline}
We initially developed our foundational protocol, CryptoTrain-B, which ingeniously merges OT~\cite{rathee2020cryptflow2, rathee2021sirnn} for non-linear operations with HE-based protocols~\cite{huang2022cheetah} for linear tasks. Notably, CryptoTrain-B substantially cuts down on training time by eliminating the intensive FHE bootstrapping required in FHE-exclusive approaches~\cite{nandakumar2019towards, lou2020glyph}.
The protocol we propose is an interactive secure training framework. At the start of the \(i\)-th layer in the neural network, the client possesses \(x_i\) and \(w_i\). Here, \(x_i\) represents the feature vector derived from processing the input \(x\) through the first \(i-1\) layers of the network (with \(x_1\) being equal to \(x\)), and \(w_i\) is the set of weights for the \(i\)-th layer. This setup is consistently maintained across each layer. We will now outline the protocol for computing the \(i\)-th layer, which involves a combination of linear and activation functions.


\noindent \textbf{Linear layer. }
Following Cheetah~\cite{huang2022cheetah}, we decompose the linear operations into inner products of polynomials. The mapping functions $\pi^w$ and $\pi^i$ are specialized for computing the inner product using polynomial arithmetic~\cite{huang2022cheetah}. The client first uses the mapping functions to transform their input $x_i$ and $w_i$ into polynomials then encrypt them. After that, the client send the ciphertexts $[x_i]$ and $[w_i]$ to the server, who will then perform a ciphertext-ciphertext multiplication (CCMul) and return the result back. 
Upon decryption by the client, the results yield the gradients of the weights during the backward propagation phase and the intermediate features in the forward propagation phase.

\begin{figure}[h!]
    \centering
    \includegraphics[width=0.95\linewidth]{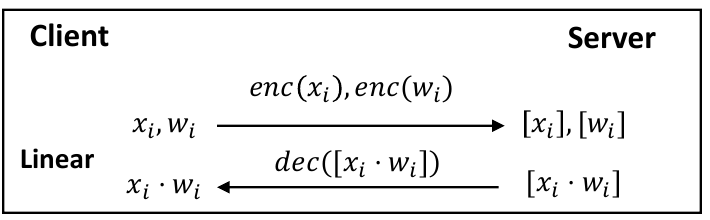}
    \caption{Linear protocol of CryptoTrain-B .}
    \label{fig:CC_linear}
\end{figure}



\begin{figure}[hb!]
    \centering
    \includegraphics[width=0.95\linewidth]{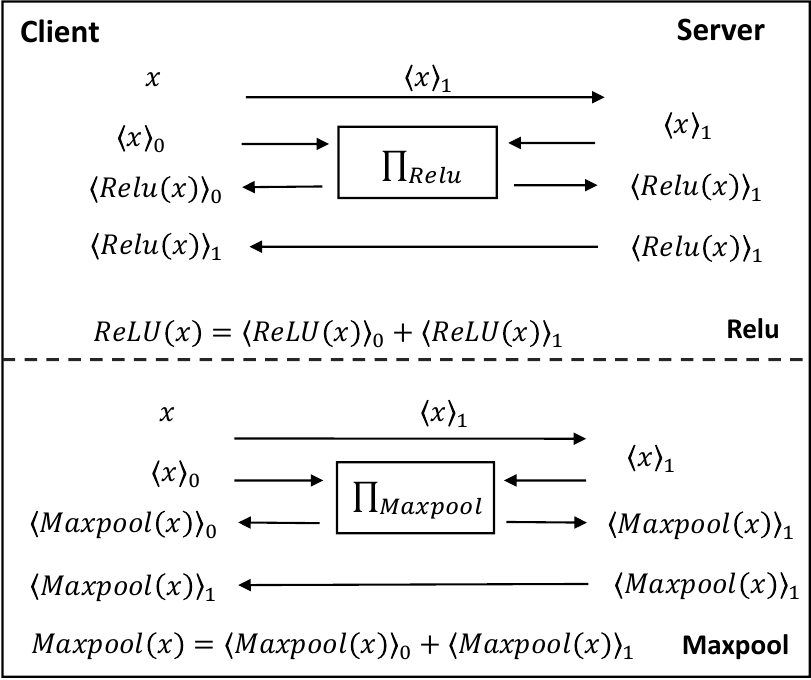}
    \caption{Non-linear protocols of CryptoTrain.}
    \label{fig:Non-linear}
\end{figure}

\noindent \textbf{Non-linear layer. }
We adopt the state-of-the-art OT-based protocols~\cite{rathee2021sirnn, rathee2020cryptflow2} to compute the non-linear functions. After the linear functions of $i$-th layer, the client is left with the plaintext $x_i$. To perform Relu and Maxpooling, the client first generates secret shares, denoted as $\langle x_i\rangle_0$ and $\langle x_i\rangle_1$, and dispatches $\langle x_i\rangle_1$ to the server, as depicted in Figure~\ref{fig:Non-linear}. After the non-linear protocols, i.e., $\prod_{Relu}$ or $\prod_{Maxpool}$ are applied, the server and client each possess a share of the result. The server then sends its share to the client who reconstructs the final result by combining both shares.

\subsection{CryptoTrain-P: CCMul-Precompute}
\label{s:CP_cryptotrain}

\begin{figure}[hb!]
    \vspace{-0.2cm}
    \centering
    \includegraphics[width=0.95\linewidth]{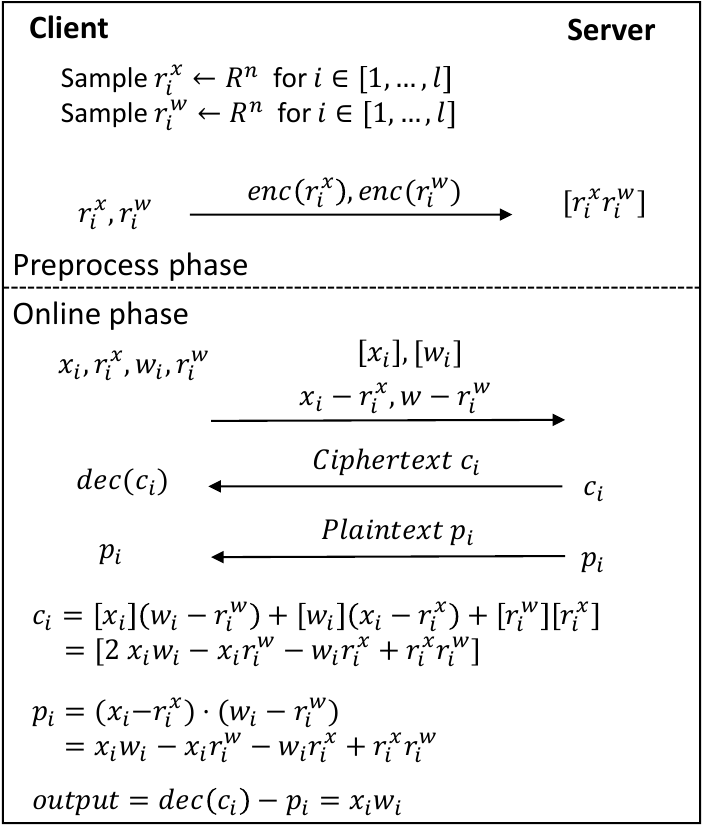}
    \caption{Linear protocol of CryptoTrain.}
    \label{fig:algorithm}
     \vspace{-0.3cm}
\end{figure}

\textcolor{black}{
As depicted in Figure ~\ref{fig:algorithm}, CryptoTrain-P splits the protocols for linear layers into an offline preprocessing phase and an online phase such that most of the heavy cryptographic computation, i.e., CCMul, is offloaded to the preprocessing phase.}

\noindent \textbf{Preprocessing Phase. }
During the preprocessing phase, the client and server engage in pre-computation of data, which is utilized later during the online execution phase. Such pre-computation is executed independently of the input values, which can be performed before the client's input and weights are known. As illustrated in the Figure~\ref{fig:algorithm}, the client first chooses two random masking vectors $r_x,r_w\xleftarrow{}R^n$. After that, the client encrypts them to $[r_i^x]$ and $[r_i^w]$, then sends them to the server. The server computes $[r_i^x][r_i^w]$ through a CCMul and holds the ciphertext result.

\noindent \textbf{Online Phase. }
At the beginning of the $i$-th layer, the client holds $x_i$, $r_i^x$, $w_i$ and $r_i^w$. Firstly, the client encodes matrices $x_i$ and $w_i$ in polynomials $\hat{x_i}$ and $\hat{w_i}$ using the same approach of CryptoTrain-B. After that, the client encrypts and sends the $[x_i]$ and $[w_i]$, as well as the secret shares $x_i-r_i^x$ and $w_i-r_i^w$ to the server. The server performs two CPMul, i.e., $c_1=[x_i]\cdot (w-r_i^w)$ and $c_2=[w_i]\cdot (x-r_i^x)$ to get $c_i$, where $c_1$ and $c_2$ are added with pre-computed $[r_i^x\cdot r_i^w]$. The server also performs a plaintext-plaintext multiplication (PPMul), i.e., $p_i=(x_i-r_i^x)\cdot (w_i-r_i^w)$. Finally, the server sends back the plaintext result $p_i$ and ciphertext result $c_i$ to the client. The client decrypts the $c_i$ and subtracts $p_i$ to get the target result $x_i\cdot w_i$. In summary, the client side reduces the need for the more costly ciphertext multiplication on the server side by performing three significantly cheaper plaintext subtractions.

\noindent \textbf{Non-linear layer. }
For non-linear protocols, CryptoTrain-P adheres to the same established protocols as those defined in the baseline configuration in our CryptoTrain-B. 

\subsection{Correlation-aware Polynomial Convolution}
\label{s:padding}

\begin{figure*}[h!]
    \centering
    \includegraphics[width=\linewidth]{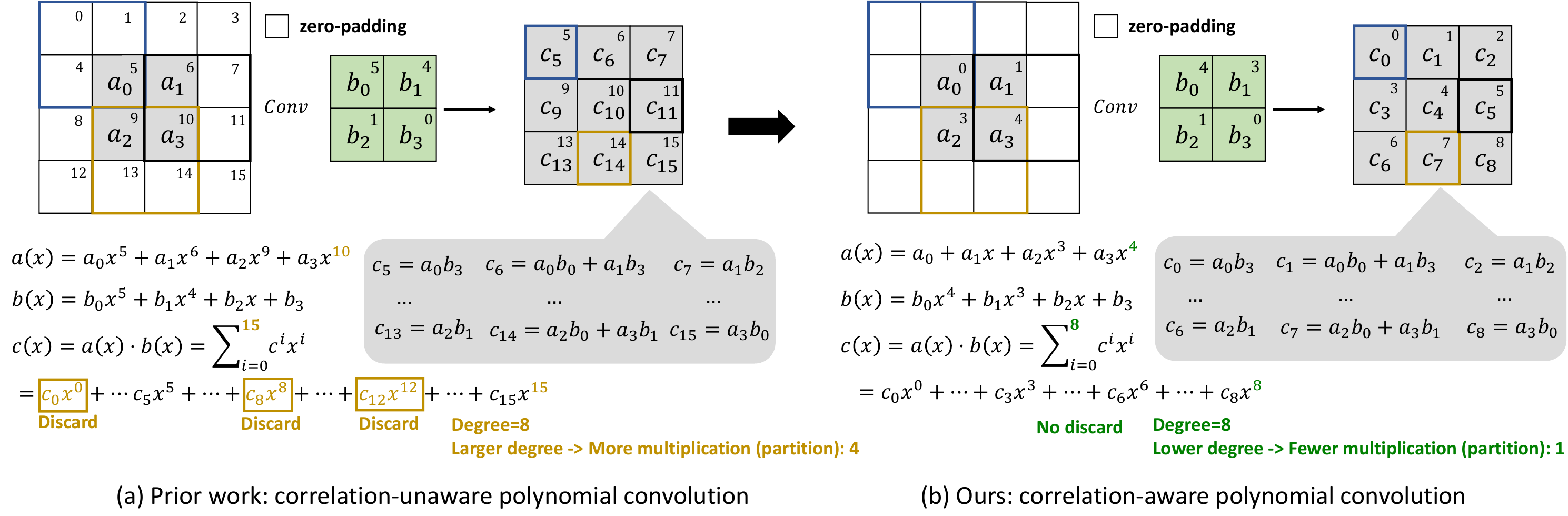}
    \caption{A toy example of our correlation-aware convolution.}
    \label{fig:Padding}
\end{figure*}
Directly adopting existing protocol to padded convolution results in an increased amount of polynomial multiplication, mainly due to the multiplication between uncorrelated terms. 
To this end, we propose a specialized correlation-aware polynomial packing scheme for padded convolution on top of Cheetah~\cite{huang2022cheetah}. By compactly packing correlated inputs and weights into polynomials in the plaintext, the proposed scheme significantly improves efficiency of polynomial encoding and reduces the number of polynomial multiplications.



Consider the example in Figure~\ref{fig:Padding}, which illustrates a convolution operation using a $2\times2$ input and a $2\times2$ kernel, with a padding size of $1$. Cheetah (left) begins by padding zero values around the input, representing it with polynomial $a(x)$. It then uses polynomial $b(x)$ for the kernel. The convolution is executed as $c(x)=a(x)\cdot b(x)$, but not all coefficients in $c(x)$ such as $c_8x^8$ and $c_{12}x^{12}$ are utilized. Consequently, when polynomial degrees are capped, i.e., $8$, Cheetah must partition a single polynomial multiplication into several so that all the polynomial degrees are in the valid range. 
In contrast, our method (right) transforms input and kernel into $a(x)$ and $b(x)$ more efficiently, reducing their degrees from $10$ and $5$ to $4$ each. Consequently, all $9$ items in the result polynomial $c(x)=a(x)\cdot b(x)$ contribute to the $3\times 3$ output. Such a higher utilization rate of coefficients, i.e., $100\%$ results in a lower polynomial degree for $c(x)$ compared to previous methods. This reduction enables the entire operation to be performed within a single polynomial multiplication, eliminating the need for partitioning. As a result, our method decreases the polynomial multiplication from $4$ to $1$.

We give the definitions of two input packing functions over the ring $\mathbb{A}_p$, $\pi_x:\mathbb{Z}^{H\times W}_p\rightarrow \mathbb{A}_{N,p}$ and $\pi_w:\mathbb{Z}^{h\times h}\rightarrow \mathbb{A}_{N,p}$ as follows:

\begin{equation}
    \hat{x} = \pi_x(x) \quad \text{where} \quad \hat{x}[iO+j] = x[i,j]
    \vspace{-0.2in}
\end{equation}

\begin{equation}
    \hat{w} = \pi_w(w) \quad \text{where} \quad \hat{w}[(h-i-1)O+(h-j-1)] = w[i,j]
\end{equation}
where $O=max(W,H)+h-1$. The multiplication $\hat{x}\cdot \hat{w}\in \mathbb{A}_{N,p}$ directly gives the $2$-dimension convolution with zero-padding. From our algorithm, we can see our padding-aware convolution does not need to assign coefficients and degrees to zero-paddings.

For large input or kernel where $W+HO>N$, we need to partition them so that each of the smaller blocks can be fit into one polynomial in $\mathbb{A}_{N,p}$. Particularly, we define the size of the partition window, $H_w$ and $W_w$. The size of the partition window can be chosen freely as long as they satisfy the following constraints: $h\leq H_w\leq H$, $h\leq W_w \leq W$, and $W_w+H_wO_w\leq N$, where $O_w=max(W_w,H_w)+h-1$. At the same time, to minimize the number of polynomial multiplications and the ciphertexts sent by the client, we choose the window size $H_w$ and $W_w$ that minimize the product $\lceil \frac{H_wW_w}{N} \rceil \cdot \lfloor \frac{H-h+1}{H_w-h+1} \rfloor \cdot \lfloor \frac{W-h+1}{W_w+h-1} \rfloor$.

\noindent \textbf{Complexity analysis. }
Consider a full-padding scenario, where the input $x\in\mathbb{Z}_p^{H\times W}$ and kernel $w\in\mathbb{Z}_p^{h\times h}$. Assuming $H=W$ for simplicity. For patching approach proposed in Cheetah~\cite{huang2022cheetah}, it needs to pad the input $x$ in to $x'\in\mathbb{Z}_p^{(H+h-1)\times(W+h-1)}$ firstly. Then encodes the $x'$ with the largest degree of $N_1=(\lfloor \frac{h-1}{2}\rfloor+H-1)\cdot (H+h-1)+H+\lfloor \frac{h-1}{2}\rfloor$. However, our method only needs to encoder $x$ into a polynomial with the largest degree of $N_2=(H+h-1)(H-1)$, which means our method needs $N_1-N_2=\lfloor \frac{h+1}{2}\rfloor H+\lfloor \frac{h-1}{2}\rfloor h$ lower degree to transform input and kernel into polynomials.

\section{Experimental Methodology} 
\noindent \textbf{Underlying Protocols.}
CryptoTrain is built on prior protocols~\cite{huang2022cheetah, rathee2021sirnn, rathee2020cryptflow2}. Specifically, we adopt the polynomial transformation strategy proposed by~\cite{huang2022cheetah} to construct efficient linear protocols for the Fully Connected (FC) layer, the Convolution layers, and the Batch-Normalization layers, which is based on HE and SS. Non-linear protocols for Relu and Maxpooling are built upon the protocol proposed in~\cite{rathee2021sirnn}, which are based on OT and ASS.

\noindent \textbf{Datasets and models.}
Our encrypted datasets include MNIST~\cite{lecun1998mnist} that has $60,000$ $28\times28$ images, CIFAR-10~\cite{krizhevsky2009learning} that has $50,000$ $32\times 32$ colored RGB images, and Tiny-ImageNet~\cite{tiny-imagenet} which contains $100,000$ images of $200$ classes downsized to $64\times64$ colored images. We consider the following training benchmark: MNIST-LeNet5~\cite{lecun1998gradient}, CIFAR-LeNet5 and TinyImageNet-AlexNet~\cite{krizhevsky2012imagenet}.

\noindent \textbf{Implementation Details.}
CryptoTrain is developed on Cheetah~\cite{huang2022cheetah} and EzPC~\cite{EzPC} frameworks, while also utilizing the SEAL cryptographic library~\cite{sealcrypto}. The EzPC framework adeptly translates deep neural networks from TensorFlow into secure computation protocols, and Cheetah underpins this with robust support for polynomial patching and multiplication processes. Consistent with methodologies in~\cite{huang2022cheetah, rathee2021sirnn}, our network simulations are set to operate at a bandwidth of $400Mbps$ and a ping latency of $0.5ms$. We conduct all our experiments on a high-performance Intel Core i9-12900F CPU, which runs at 2.4GHz and is equipped with 64GB DRAM, boasting $16$ cores and the capability to support $24$ threads.

\section{Results and Analysis}

\noindent \textbf{Comparison to existing protocols.}
In Table~\ref{t:vs_baseline_ot}, we evaluate CryptoTrain across two models and three datasets compared with our baseline and the state-of-the-art MPC-based secure training, KS22~\cite{keller2022secure} and HE-based method Glyph~\cite{lou2020glyph}. CryptoTrain demonstrates a speedup of up to $8.2\times$ compared to the baseline in the TinyImageNet-AlexNet setting, and speedups of $2.2\times$ and $2.3\times$ on MNIST-LeNet5 and CIFAR-LeNet5 settings, respectively. 
When compared to KS22 and Glyph, CryptoTrain achieves speedups of $5.3\times$ and $75.5\times$, respectively. These gains are attributed to our optimized polynomial multiplication, the shifting of intensive computations to the preprocessing phase, and the adoption of OT-based protocols for non-linear operations. Although the communication between client and server increase since the client needs to share plaintexts \(x_i-r_i^x\) and \(w-r_i^w\) to server, such additional communication enables the avoidance of CCMul operations, leading to a significant reduce of latency.



\begin{table}[ht!]
\centering
\footnotesize
\setlength{\tabcolsep}{3.1pt}
\captionsetup{skip=2pt}
\caption{Comparison on Time and Communication.}
\begin{tabular}{ccccccc}\toprule
Model & Dataset & Method & Epoch & ACC(\%) & T(h) & Comm.(MB)\\\midrule
\multirow{6}{*}{LeNet-5} & \multirow{3}{*}{MNIST} & Glyph~\cite{lou2020glyph} & $10$ & $99.23$ & $576$ & $-$\\
 & & CryptoTrain-B & $10$ & $99.23$ & $16.11$ & $51.41$ \\ 
 & & CryptoTrain & $10$ & $99.23$ & $7.36$ & $149.4$ \\\cmidrule(lr){2-7}
 & \multirow{3}{*}{CIFAR-10} & KS22~\cite{keller2022secure} & $15$ & $67.64$ & $83.7$ & $134.6$\\  
 & &  CryptoTrain-B & $15$ & $67.64$ & $32.12$ & $55.2$ \\ 
 & & CryptoTrain & $15$ & $67.64$ & $15.72$ & $170.4$\\\midrule
\multirow{2}{*}{AlexNet} & \multirow{2}{*}{Tiny-ImageNet} &  CryptoTrain-B & $25$ & $30.82$ & $4351$ & $832$ \\ 
 & & CryptoTrain & $25$ & $30.82$ & $528.1$ & $1475$ \\
\bottomrule
\end{tabular}
\vspace{-0.3cm}
\label{t:vs_baseline_ot}
\end{table}

\noindent \textbf{Ablation study on different techniques. }
In Table~\ref{t:ablation}, we evaluate the effectiveness of the techniques we proposed on TinyImageNet-AlexNet setting, focusing on latency and communication overhead for a single epoch. The results indicate a substantial reduction in latency, from $174.0$ to $49.2$ hours, when offloading CCMul to the preprocessing phase, albeit at a $3.5\times$ increase in communication costs. Additionally, the implementation of our correlation-aware polynomial convolution further reduces the run-time to $21.1$ hours and cuts the communication costs by approximately half. This efficiency gain is attributed to our method's requirement for fewer partitions when performing convolution with padding, thereby decreasing the amount of ciphertexts client need to transmit and consequently lowering the number of communications and computations.


\begin{table}[th!]
\centering
\footnotesize
\captionsetup{skip=2pt}
\caption{Ablation study on TinyImageNet-AlexNet setting.}
\begin{tabular}{ccc}\toprule
Protocol & Time (Hour) & Comm. (MB) \\\midrule
CryptoTrain-B & $174.0$ & $832$  \\
CryptoTrain-P & $49.2$ & $2901$ \\
CryptoTrain & $21.1$ & $1473$ \\
\bottomrule
\end{tabular}
\vspace{-0.1in}
\label{t:ablation}
\end{table}

\begin{figure}[ht!]
    \centering
    \includegraphics[width=\linewidth]{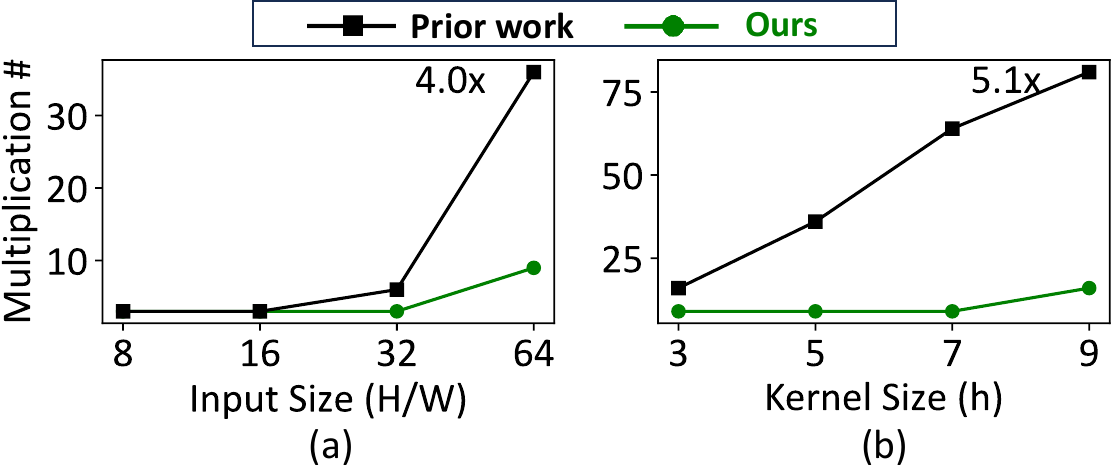}
    \caption{Multiplication Counts in Padded Convolution.}
    \label{fig:padding_ablation}
\end{figure}

\noindent \textbf{Ablation study on kernel size and input size.}
Moreover, we compare our correlation-aware polynomial convolution with prior work~\cite{huang2022cheetah} across a range of input and kernel sizes. As illustrated in Figure~\ref{fig:padding_ablation}(a), we apply a $5\times 5$ kernel to inputs varying from $8\times8$ to $64\times64$, maintaining both input and output channels at $3$ for simplicity. The results demonstrate that our convolution scheme yields more substantial performance enhancements with larger input sizes, which is also reflected in Table~\ref{t:vs_baseline_ot}. Specifically, our method can achieve a $4$-fold reduction in multiplication operations compared to prior work. Additionally, we experimented with different kernel sizes while keeping the input size constant at $64\times64$. Figure~\ref{fig:padding_ablation}(b) reveals that our method surpasses previous approaches, delivering even more pronounced benefits for larger kernels. For example, our scheme can decrease the number of multiplications by a factor of $5.1$.

\noindent \textbf{Performance breakdown. }
In Table~\ref{t:breakdown}, we study the performance of CryptoTrain on main operations of AlexNet: Convolution, Relu, MaxPooling, Batch Normalization, and FC, corresponding to the initial three layers and the first fully connected layer, respectively. The results highlight CryptoTrain's notable improvements in latency for linear operations. For the FC layer, the latency decreases from $533.9$ ms to $145.1$ ms due to the offloading of CCMul. For the Convolution layer, the latency even drops more from $2,913$ to $464.9$ ms because of our correlation-aware polynomial convolution.

\begin{table}[th!]
\centering
\footnotesize
\captionsetup{skip=1pt}
\caption{Overhead of different layers.}
\begin{tabular}{ccccccc}\toprule
\multirow{2}{*}{Method} & \multirow{2}{*}{Overhead} & \multicolumn{3}{c}{Linear} & \multicolumn{2}{c}{Non-linear}\\\cmidrule(lr){3-5} \cmidrule(lr){6-7}
& & Conv & FC & BN & Relu & Maxpool\\\midrule
\multirow{2}{*}{CryptoTrain-B} & Latency (ms) & $2913$ & $533.9$ & $71.34$ & $20.17$ & $124.6$\\
& Comm.(MB) & $10.8$ & $7.63$ & $2.38$ & $2.11$ & $29.30$\\\midrule
\multirow{2}{*}{CryptoTrain} & Latency (ms) & $464.9$ & $145.1$ & $21.0$ & $20.17$ & $124.6$\\
& Comm.(MB) & $16.7$ & $22.61$ & $24.38$ & $2.11$ & $29.30$\\\bottomrule
\label{t:breakdown}
\end{tabular}
\vspace{-0.2in}
\end{table}

\noindent \textbf{End-to-End Performance.}
To quantitatively assess the preprocessing phase duration within CryptoTrain, we executed experiments to measure the end-to-end time, encompassing both preprocessing and online phases, as detailed in Table~\ref{t:end-to-end}. The results show that the total time for CryptoTrain in the AlexNet-TinyImageNet configuration amounts to $1704$ hours. Remarkably, this end-to-end training time is $2.6 \times$ shorter than that observed with our baseline protocol. Such a significant reduction in end-to-end time underscores the effectiveness of our correlation-aware polynomial convolution, notably in reducing CCMul operations required during the preprocessing phase. This result not only highlights CryptoTrain's efficiency but also validates the impact of our optimization strategies on the overall training process.

\begin{table}[th!]
\centering
\footnotesize
\captionsetup{skip=2pt}
\caption{End-to-end time (in hours) and communication (MB).}
\begin{tabular}{cccccc}\toprule
Settings  & Method & Time & Comm. & Epoch & ACC(\%)\\\midrule
\multirow{2}{*}{\makecell{LeNet5-\\CIFAR10}} & CrytoTrain-B & $32.12$ & $55.2$ & \multirow{2}{*}{$15$} & \multirow{2}{*}{$67.64$} \\
 & CrytoTrain & $41.69$ & $170.4$ & & \\\midrule
\multirow{2}{*}{\makecell{AlexNet-\\TinyImageNet}}  & CrytoTrain-B & $4351$ & $832$ & \multirow{2}{*}{$25$} & \multirow{2}{*}{$30.82$} \\
& CrytoTrain & $1704$ & $1475$ & &  \\\bottomrule 
\end{tabular}
\vspace{-0.5cm}
\label{t:end-to-end}
\end{table}

\section{Conclusion}
In this paper, we introduce CryptoTrain, an efficient framework for secure training. CryptoTrain, built on a hybrid cryptographic protocol, strategically offloads heavy ciphertext-ciphertext multiplications (CCMul) to a preprocessing phase, reducing latency during training. Additionally, our innovative correlation-aware polynomial convolution enhances the efficiency of padded convolutions. CryptoTrain establishes a new framework for efficient secure training, eliminating the need for non-colluding servers and advancing the field significantly.



\section{Discussion}
Our work on secure training protocols presents several advancements in the field of Training as a Service (TaaS), yet there remain areas for potential optimization and further research. A crucial direction for improvement is reducing the communication overhead between the client and the server. Recent research~\cite{mahdavi2023he} has proposed a method for it. While traditional Homomorphic Encryption schemes typically encrypt a plaintext message into two polynomials that are both sent to the server, \cite{mahdavi2023he} proposes compressing one of these ciphertext polynomials into a Pseudorandom Generator (PRG) seed. This approach allows the client to send only the seed and the other polynomial to the server, which can then reconstruct the compressed polynomial. By effectively reducing the communication between server and client by almost half, this technique significantly improves the efficiency of data transmission in secure training protocols, potentially making TaaS more practical and scalable, especially in bandwidth-constrained environments.

\bibliography{main}
\balance
\bibliographystyle{ACM-Reference-Format}
\appendix

\end{document}